\documentclass[
    ,final            
  ]
  {aipproc}

\layoutstyle{6x9}

\usepackage{amsmath}
\usepackage{amssymb}
\usepackage{mathrsfs}

\newcommand{\Msun}{\:{\rm M}_{\odot}}
\newcommand{\Mtwohun}{\:{\rm M}_{\rm halo}}
\newcommand{\rhalf}{{\rm r}_{_{1/2}}}
\newcommand{\Rhalf}{{\rm R}_{_{\rm e}}}

\newcommand{\Mhalf}{{\rm M}_{_{1/2}}}
\newcommand{\Lhalf}{{\rm L}_{_{1/2}}}
\newcommand{\moverl}{\Upsilon^{\, \rm I}_{_{1/2}}}
\newcommand{\moverlI}{\Upsilon^{\, \rm I}_{_{1/2}}}
\newcommand{\sigmalos}{\sigma_{_{\rm{los}}}}
\newcommand{\Lsun}{{\rm L}_\odot}

\newcommand{\LI}{{\rm L}_{\rm I}}
\newcommand{\LV}{{\rm L}_{\rm V}}

\newcommand{\beq}{\begin{equation}}
\newcommand{\eeq}{\end{equation}}

\newcommand{\avelos}{\langle \sigma_{_{\rm los}}^2 \rangle}

\begin{document}

\title{Testing galaxy formation scenarios with a new mass estimator}


\author{Joe Wolf}{
  address={Center for Cosmology, Department of Physics and Astronomy \\ University of California, Irvine, CA 92697; wolfj@uci.edu}
}

\begin{abstract}
We present the recently derived Wolf et al. (2009) mass estimator, which
is applicable for spherical pressure-supported stellar systems spanning
over ten orders of magnitude in luminosity, as a tool to test galaxy formation theories.
We show that all of the Milky Way dwarf spheroidal galaxies (MW dSphs) are consistent 
with having formed within a halo of mass approximately $3 \times 10^9 \Msun$ in
$\Lambda$CDM cosmology. The faintest MW dSphs seem to have formed in dark
matter halos that are at least as massive as those of the brightest
MW dSphs, despite the almost five orders of magnitude spread in luminosity.
We expand our analysis to the full range of observed
pressure-supported stellar systems and examine their I-band
mass-to-light ratios $\moverlI$. The $\moverlI$ vs. half-light mass $\Mhalf$ relation for 
pressure-supported galaxies follows a U-shape, with a broad minimum 
near $\moverlI \simeq 3$ that spans dwarf elliptical galaxies to normal 
ellipticals, a steep rise to $\moverlI \simeq 3,200$ for ultra-faint dSphs, 
and a more shallow rise to $\moverlI \simeq 800$ for galaxy cluster spheroids.  
\end{abstract}

\maketitle

\section{Accurate mass determinations from L.O.S. kinematics}

Mass determinations for pressure-supported galaxies based on only line-of-sight velocity measurements suffer from a notorious uncertainty associated with not knowing the intrinsic 3D velocity dispersion. The difference between radial and tangential velocity dispersions is usually quantified by the stellar velocity dispersion anisotropy, $\beta$. Many questions in galaxy formation are affected by our ignorance of $\beta$, including our ability to quantify the amount of dark matter in the outer parts of elliptical galaxies, to measure the mass profile of the Milky Way from stellar halo kinematics, and to infer accurate mass distributions in dwarf spheroidal galaxies (dSphs).

However, it has recently been shown that by manipulating the spherical Jeans equation 
there exists one radius within any pressure-supported galaxy where the
integrated mass as inferred from the line-of-sight velocity dispersion
is largely insensitive to $\beta$, and that this radius is
approximately equal to the 3D deprojected half-light radius $\rhalf$ \cite{Wolf_09}.
Moreover, the mass within $\rhalf$ is well characterized by a simple
formula that depends only on quantities that may be inferred from
observations:
\beq
\label{eq:main}
\Mhalf \equiv M(\rhalf) \simeq 3 \, G^{-1} \, \avelos\, \rhalf \simeq 
4 \, G^{-1} \, \avelos \, \Rhalf  \simeq 
930 \: \left(\frac{\avelos}{\rm km^2 \, s^{-2}} \right)
 \: \left(\frac{\Rhalf}{\rm pc}\right) \: \Msun\,,
\eeq
where $M(r)$ is the mass enclosed within a sphere of radius $r$, $\sigmalos$
is the line-of-sight velocity dispersion and the brackets indicate 
a luminosity-weighted average. In the above equation we have used $\Rhalf \simeq (3/4) \,
\rhalf$ for the 2D projected half-light radius. This approximation is
accurate to better than 2\% for exponential, Gaussian, King, Plummer,
and S\'ersic profiles (see Appendix B of \cite{Wolf_09}).
%
%
\begin{figure*}
\includegraphics[width=68mm]{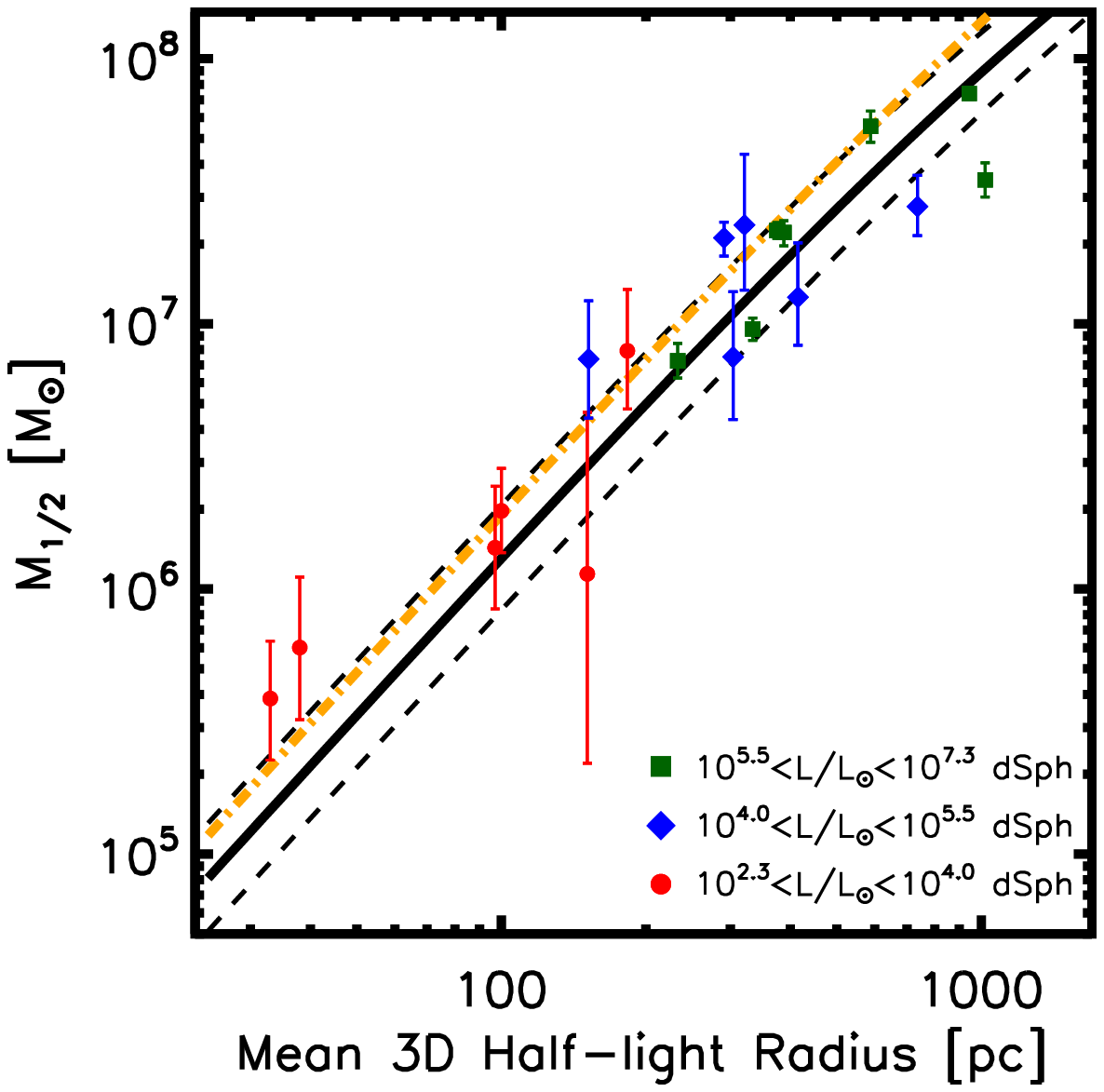}
\hspace{5mm}
\includegraphics[width=68mm]{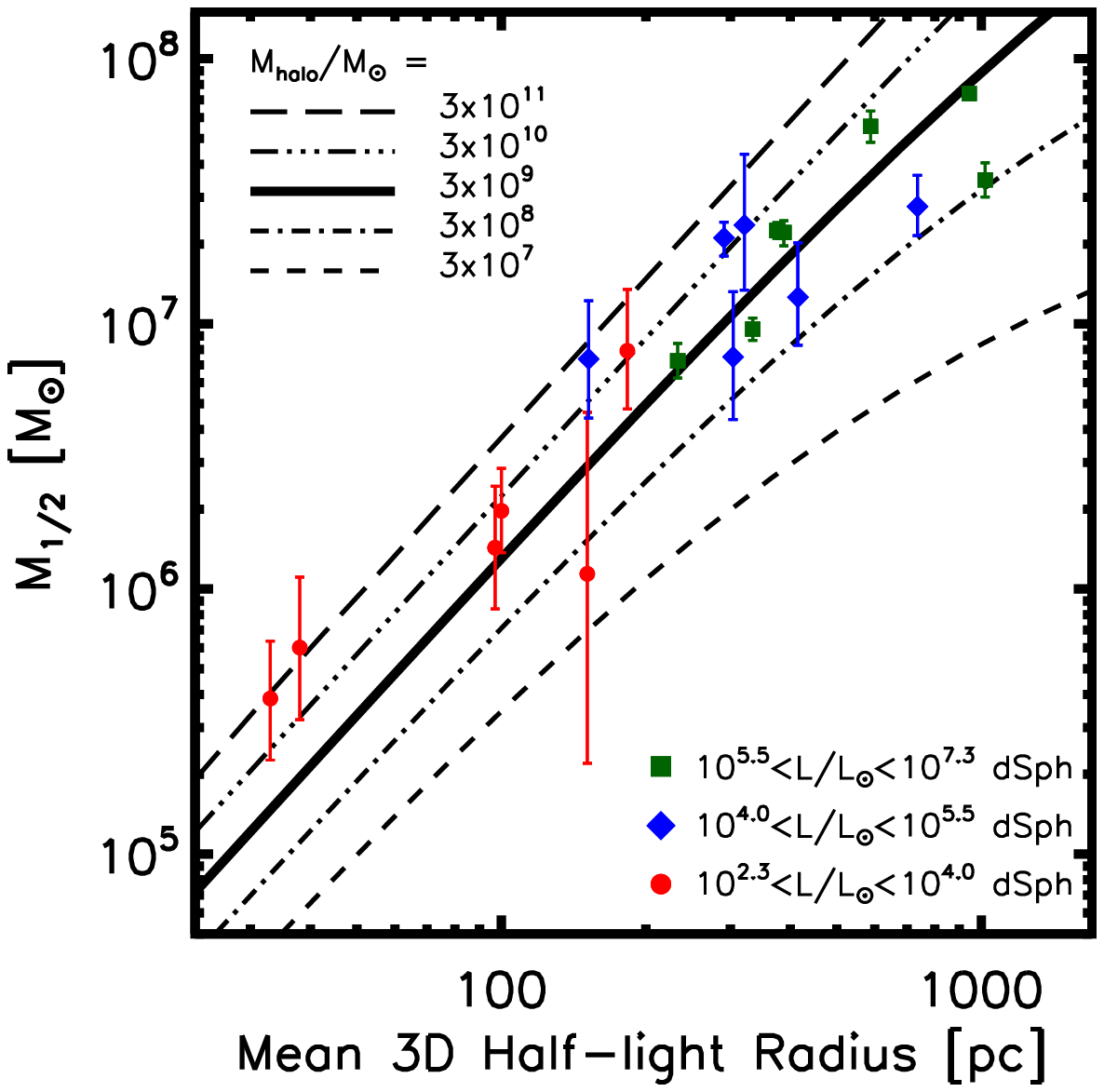}
\caption{The half-light masses of the Milky Way dSphs plotted against $\rhalf$. See text for description.}
\label{fig:Mrhalfvsrhalf}
\end{figure*} 
%
%
\section{Dwarf spheroidal galaxies of the Milky Way}

As an example of the utility of $\Mhalf$ determinations, 
Figure \ref{fig:Mrhalfvsrhalf} presents $\Mhalf$ vs. $\rhalf$ for MW
dSph galaxies. Relevant parameters for each of the galaxies
are provided in Table 1 of \cite{Wolf_09}. The symbol types labeled on the plot
correspond to three wide luminosity bins that span almost five orders
of magnitude in luminosity. It is interesting now to compare the data points in Figure
\ref{fig:Mrhalfvsrhalf} to the integrated mass profile $M(r)$
predicted for $\Lambda$CDM dark matter field halos of a given $\Mtwohun$
mass, which is defined as the halo mass corresponding to an
overdensity of 200 compared to the critical density. In
the limit that dark matter halo mass profiles $M(r)$ map 
in a one-to-one way with their $\Mtwohun$ mass, then the
points on this figure may be used to estimate an associated halo mass
for each galaxy. 

The solid line in the left panel of Figure \ref{fig:Mrhalfvsrhalf}
shows the mass profile for a NFW \citep{nfw} dark matter halo at
$z=0$ with a halo mass  $\Mtwohun = 3 \times 10^9 \Msun$. We have used
the median concentration ($c=11$) predicted by the mass-concentration model 
\cite{Bullock_01} updated for WMAP5 $\Lambda$CDM parameters \cite{Maccio_08}.
The dashed lines
indicate the expected $68 \%$ scatter about the median concentration
at this mass. The dot-dashed line shows the expected $M(r)$ profile
for the same mass halo at $z=3$ (corresponding to a concentration of $c=4$), 
which provides an estimate of the scatter that would result from the 
scatter in infall times. We see that each MW dSph is consistent with 
inhabiting a dark matter halo of mass $\sim 3 \times 10^9 \Msun$
\cite{Strigari_08}.

The right panel in Figure \ref{fig:Mrhalfvsrhalf} shows the same data
plotted along with the median mass profiles for several different halo
masses. Clearly, the data are also consistent with MW dSphs populating
dark matter halos of a wide range in $\Mtwohun$ above $\sim 3 \times 10^8 \Msun$.
This value provides a very stringent limit on the fraction of the baryons converted to stars in these
systems. More importantly, there is no systematic relationship
between dSph luminosity and the $\Mtwohun$  mass profile that they
most closely intersect. The ultra-faint dSph population (circles)
with $\LV < 10,000 \, \Lsun$ is equally likely to be associated with
the more massive dark matter halos as are classical dSphs that are
more than 1,000 times brighter (squares).  Indeed, a naive
interpretation of the right-hand panel of Figure
\ref{fig:Mrhalfvsrhalf} shows that the two least luminous satellites
(which also have the smallest $\Mhalf$ and $\rhalf$ values) are
associated with halos that are {\em more massive} than any of the
classical MW dSphs (squares). This general behavior is difficult
to reproduce in models constructed to confront the Milky Way satellite
population, which typically predict a noticeable trend
between halo infall mass and dSph luminosity. It is possible that we
are seeing evidence for a new scale in galaxy formation \cite{Strigari_08}
or that there is a systematic bias that makes less luminous
galaxies that sit within low-mass halos more difficult to detect than
their more massive counterparts (Bullock et al., in preparation). 
\section{Global population of pressure-supported systems}
\begin{figure*}
\includegraphics[width=46mm]{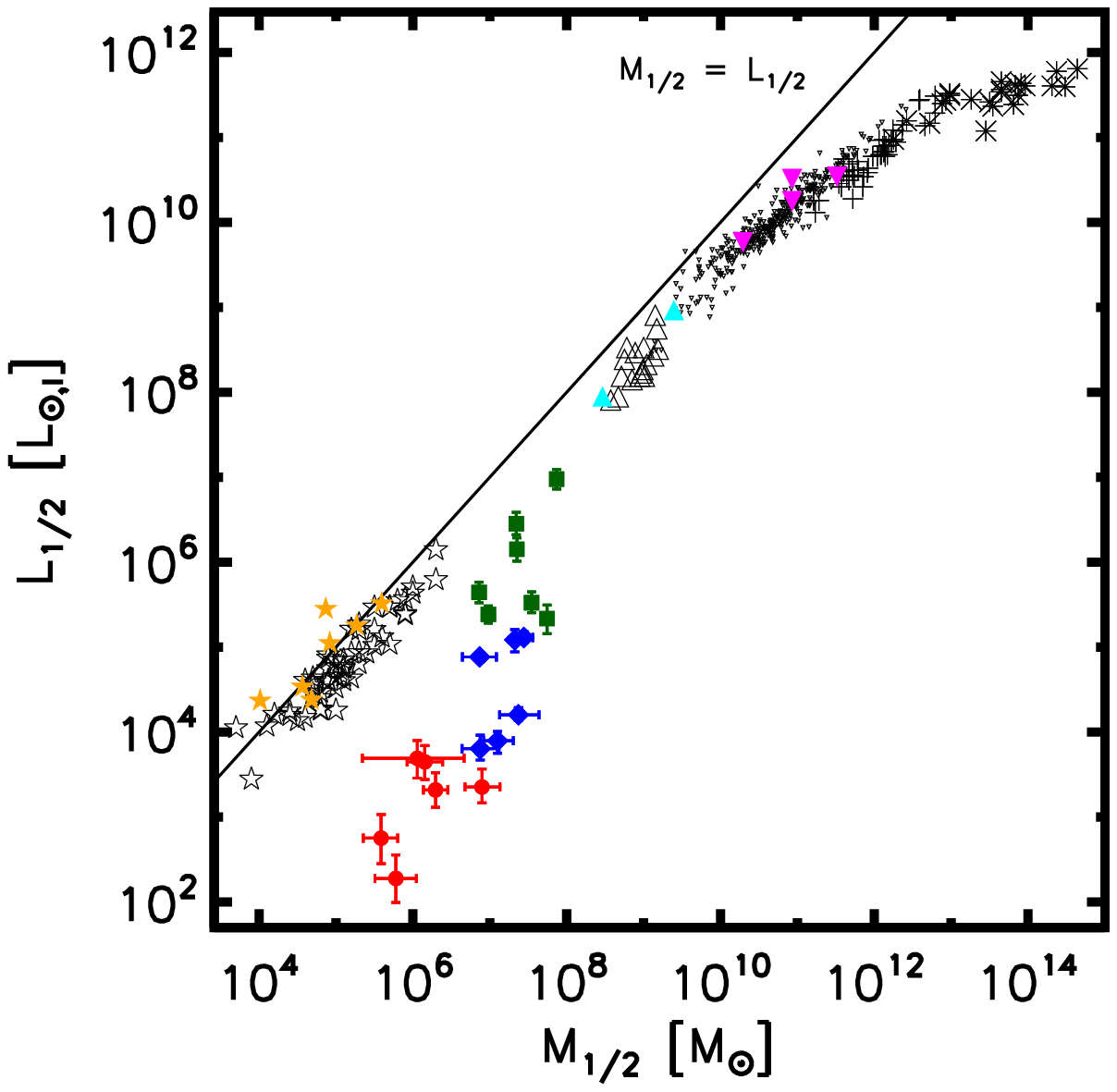}
\hspace{1mm}
\includegraphics[width=46mm]{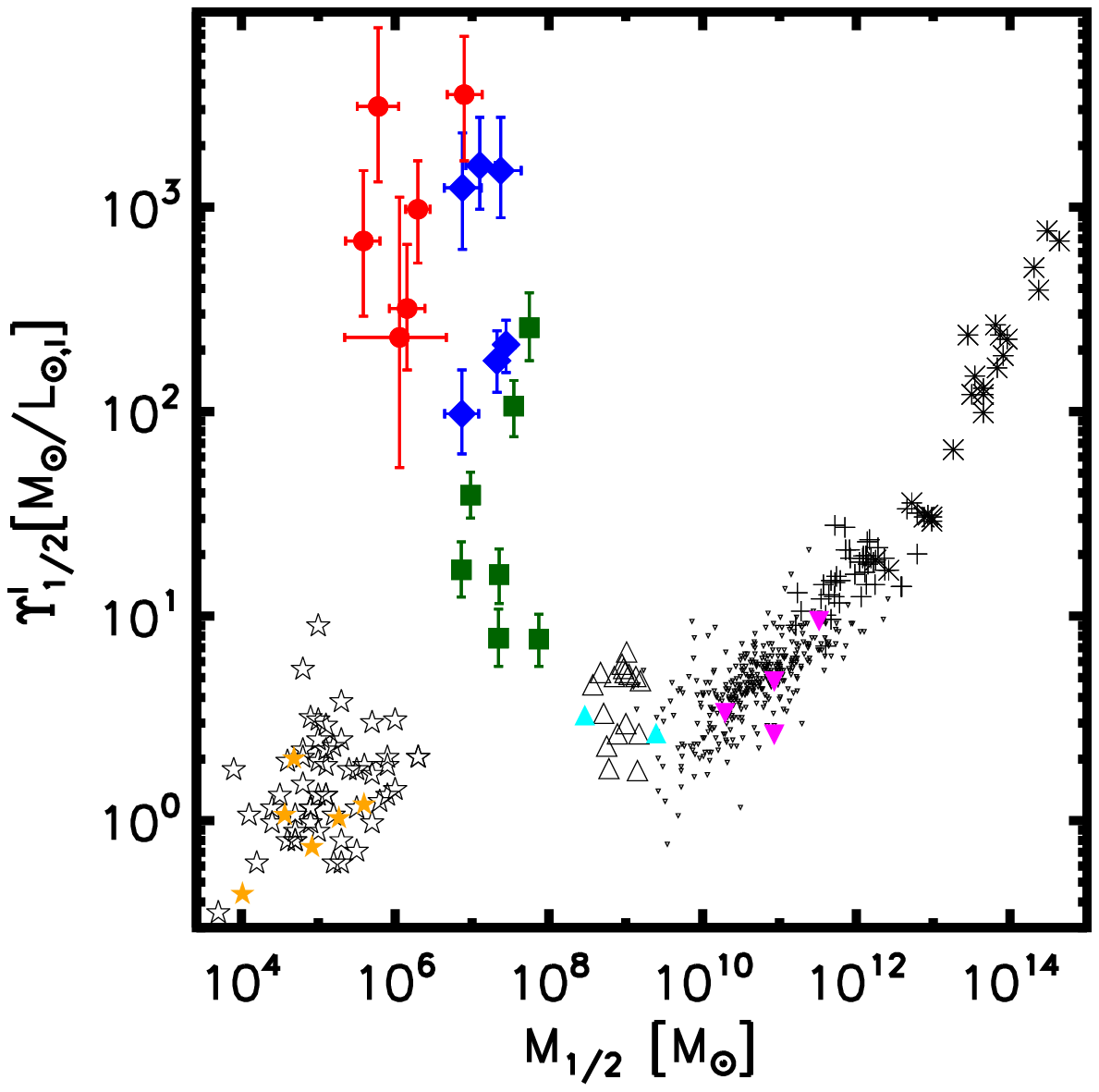}
\hspace{1mm}
\includegraphics[width=46mm]{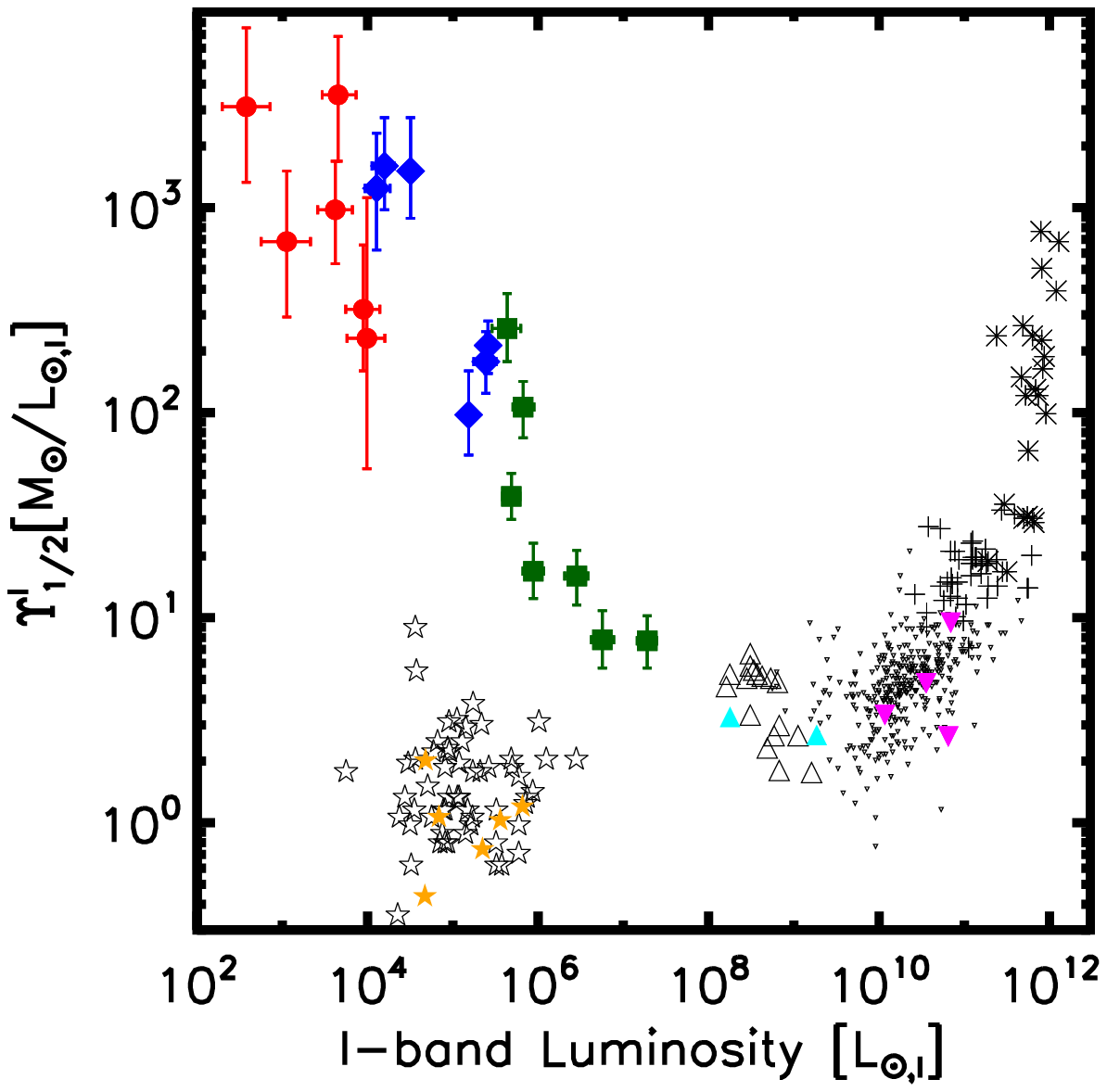}
\caption{Left: The half I-band luminosity $\Lhalf$ vs. half-light mass $\Mhalf$ for a broad population of spheroidal galaxies. Middle: The half-light mass-to-light ratio $\moverl$ vs. $\Mhalf$ relation. Right: The equivalent $\moverl$ vs. total I-band luminosity L$ = 2 \, \Lhalf$ relation. The line in the left panel guides the eye with $\Mhalf = \Lhalf$ in solar units. The symbols are linked to galaxy types as follows: Milky Way dSphs (squares, diamonds, circles), galactic globular clusters (stars), dwarf ellipticals (triangles), ellipticals (inverted triangles), brightest cluster galaxies (plus signs), and cluster spheroids (asterisks). See Figure 4 of \cite{Wolf_09} for references.}
\label{fig:manifold}
\end{figure*}

A second example of how accurate $\Mhalf$ determinations may be used
to constrain galaxy formation scenarios is presented in Figure
\ref{fig:manifold}, where we examine the relationship between the
half-light mass $\Mhalf$ and the half-light luminosity $\Lhalf = 0.5
\, \LI$ for the full range of pressure-supported stellar systems in the
universe: globular clusters, dSphs, dwarf ellipticals, ellipticals,
brightest cluster galaxies, and extended cluster spheroids.

There are several noteworthy aspects to Figure \ref{fig:manifold}, which are each highlighted in a slightly different fashion in the three panels. First, as seen most clearly in the middle and right panels, the half-light mass-to-light ratios of spheroidal galaxies in the universe demonstrate a minimum at $\moverl \simeq 2-4$ that spans a remarkably broad range of masses $\Mhalf \simeq 10^{7-10.5} \Msun$ and luminosities $\LI \simeq 10^{6.5-10} \, \Lsun$. It is interesting to note the offset in the average mass-to-light ratios between globular clusters and L$_\star$ ellipticals, which may suggest that even within $\rhalf$, dark matter may constitute the majority of the mass content of L$_\star$ elliptical galaxies. Nevertheless, it seems that dark matter plays a clearly dominant dynamical role ($\moverl \gtrsim 5$) within $\rhalf$ in only the most extreme systems. The dramatic increase in half-light mass-to-light ratios at both smaller and larger mass and luminosity scales is indicative of a decrease in galaxy formation efficiency in the smallest and largest dark matter halos. It is worth mentioning that a similar trend in the $\Mtwohun$ vs. $L$ relationship must exist if $\Lambda$CDM is to explain the luminosity function of galaxies. While the relationship presented in Figure \ref{fig:manifold} focuses on a different mass variable, the similarity in the two relationships is striking, and generally encouraging for the theory.

One may gain some qualitative insight into the physical processes that drive galaxy formation inefficiency in faint vs. bright systems by considering the $\Mhalf$ vs. $\Lhalf$ relation (left panel) in more detail.  We observe three distinct power-law regimes $\Mhalf \propto \Lhalf^\alpha$ with $\alpha <1$, $\alpha \simeq 1$, and $\alpha > 1$ as mass increases. Over the broad middle range of galaxy masses, $\Mhalf \simeq 10^{7-10.5} \Msun$, mass and light track each other quite closely with $\alpha \simeq 1$, while very faint galaxies obey $\alpha \simeq 1/2$, and bright elliptical galaxies have $\alpha \simeq 4/3$ transitioning to $\alpha \gg 1$ for the most luminous cluster spheroids. One may interpret the transition from $\alpha < 1$ in faint galaxies to $\alpha > 1$ in bright galaxies as a transition between mass-suppressed galaxy formation to luminosity-suppressed galaxy formation. That is, for faint galaxies ($\alpha < 1$), we do not see any evidence for a low-luminosity threshold in galaxy formation, but rather we are seeing behavior closer to a threshold (minimum) mass with variable luminosity. For brighter spheroids with $\alpha > 1$, the increased mass-to-light ratios are driven more by increasing the mass at fixed luminosity, suggestive of a maximum luminosity scale \cite{Wolf_09}.

Regardless of the interpretation of Figure \ref{fig:manifold}, it provides a useful empirical benchmark against which theoretical models can compare. Interestingly, two of the least luminous MW dSphs have the highest mass-to-light ratios $\moverl \simeq 3,200$ of any collapsed structures shown, including intra-cluster light spheroids, which reach values of $\moverl \simeq 800$. Not only are the ultra-faint dSphs the most dark matter dominated objects known, as they have even lower baryon-to-dark matter fractions $f_{b} \sim \Omega_{b}/\Omega_{dm} \lesssim 10^{-3}$ than galaxy clusters $f_{b} \simeq 0.1$, we now see that ultra-faint dSphs also have higher mass-to-visible light ratios within their stellar extents than even the (well-studied) galaxy cluster spheroids.
%
%

\end{document}